\documentclass[conference]{IEEEtran}
\IEEEoverridecommandlockouts

\usepackage[linesnumbered,ruled,vlined]{algorithm2e}
\DontPrintSemicolon

\SetKwComment{Comment}{\color{green!50!black}// }{}

\newcommand{\assign}{\scriptsize\leftarrow}
\newcommand{\var}[1]{\texttt{\scriptsize #1}}
\newcommand{\In}[1]{\textnormal{\scriptsize\textcolor{blue!90!black}{\ttfamily\bfseries #1}}\unskip}
\newcommand{\FuncCall}[2]{\texttt{\scriptsize\bfseries #1(#2)}}
\SetKwProg{Function}{function}{}{}
\SetKwProg{Class}{class}{}{}
\SetKwProg{For}{for}{}{}

\usepackage{cite}
\usepackage{amsmath,amssymb,amsfonts}
\usepackage{algorithmic}
\usepackage{graphicx}
\usepackage{textcomp}
\usepackage{comment}
\usepackage{xcolor}
\def\BibTeX{{\rm B\kern-.05em{\sc i\kern-.025em b}\kern-.08em
    T\kern-.1667em\lower.7ex\hbox{E}\kern-.125emX}}
\begin{document}

\title{Proof-of-Collaborative-Learning: A Multi-winner Federated Learning Consensus Algorithm}


\author{\IEEEauthorblockN{Amirreza Sokhankhosh}
\IEEEauthorblockA{\textit{University of Manitoba, Winnipeg, Canada}\\
sokhanka@myumanitoba.ca}
\and
\IEEEauthorblockN{Sara Rouhani}
\IEEEauthorblockA{\textit{University of Manitoba, Winnipeg, Canada}\\
sara.rouhani@umanitoba.ca}
}

\maketitle

\begin{abstract}
Regardless of their variations, blockchains require a consensus mechanism to validate transactions, supervise added blocks, maintain network security, synchronize the network state, and distribute incentives. Proof-of-Work (PoW), one of the most influential implementations of consensus mechanisms, consumes an extraordinary amount of energy for a task that lacks direct productive output. In this paper, we propose Proof-of-Collaborative-Learning (PoCL), a multi-winner federated learning validated consensus mechanism that redirects the computation power of blockchains to train federated learning models. In addition, we present a novel evaluation mechanism to ensure the efficiency of the locally trained models of miners. We evaluated the security of our evaluation mechanism by introducing and conducting probable attacks. Moreover, we present a novel reward distribution mechanism to incentivize winning miners fairly, and demonstrate that our reward system is fair both within and across all rounds.
\end{abstract}

\begin{IEEEkeywords}
Blockchain, Consensus, Federated Learning, Incentive Mechanism, Security, Fairness. 
\end{IEEEkeywords}

\section{Introduction} 

The decentralized nature of blockchain technology, the incentive mechanisms supporting blockchain networks, and smart contracts—which enable programmable and automated transactions—have all contributed to blockchain's explosive expansion. The Proof-of-Work (PoW) consensus mechanism, which rewards miners for their computing efforts to maintain the network, requires substantial computational power to preserve data integrity and security. The most well-known use of PoW, Bitcoin mining, has an energy demand that is currently equal to the yearly energy usage of nations like Poland \cite{digiconomist}. These serious environmental issues emphasize the need for more effective consensus mechanisms to retain blockchain's advantages while mitigating its environmental impact.

The PoW mechanism tasks miners by finding a specific nonce value via trial and error and rewarding the first successful miner. Because of the inefficiency of this brute-force approach, proposals have emerged to replace it with a more meaningful puzzle \cite{Qu2021pofl, Shoker2017pox, dong2019proofware, king2013primecoin}. Other studies \cite{nguyen2019pos, vericonomy} have also suggested alternatives with less computational requirements, using various consensus models that focus on efficiency and lower energy consumption.

Federated learning (FL), first proposed by McMahan et al. \cite{mcmahan2017communication}, is a collaborative machine learning approach that trains models across distributed networks. FL improves privacy by retaining data on individual client devices, sharing only model updates with the central server. However, this approach still raise privacy concerns for both the server and clients within an FL network \cite{zhu2023blf}. An incentive mechanism can motivate clients to truthfully contribute to the global model, thereby strengthening the server's security. In addition, a decentralized governance scheme can eradicate the security concerns of the clients. Hence, to address these problems, Kim et al. \cite{kim2020blockfl} proposed a blockchain-enabled FL system where model updates are verified and communicated using blockchain and smart contracts. Since then, numerous studies have expanded on this concept, applying it to various technological domains \cite{lu2020IoT, zhang2021IoT, Qu2021cognitive, kumar2021ctimaging, rahman2020secure, cui2022creat, pokhrel2020auto}.

In Proof-of-Federated-Learning (PoFL) \cite{Qu2021pofl}, requesters send Deep Learning (DL) tasks along with a list of possible data providers. These tasks are distributed among distinct pools of miners, with each pool selecting a DL model for the task. In each pool, a miner is selected as the leader and acts as the central server in the traditional FL paradigm, while others operate as clients. However, there are two major drawbacks to this approach: i) FL is only achieved in pools with a limited number of miners, thereby undermining the efficiency of models trained, and ii) the framework lacks fairness because the winning global model is not shared with other pools. This leads to an unfair advantage for pools that win in the initial rounds. This disparity significantly affects miners who join the competition in the middle or final rounds.

Accordingly, this paper presents Proof-of-Collaborative-Learning (PoCL), a novel decentralized multi-winner FL consensus mechanism that improves model evaluation by using a distributed network of miners. In our framework, miners distribute unlabeled test records to evaluate the trained local models of other miners, who predict these records and report their results. These predictions are evaluated based on accuracy (loss value) and timeliness (prediction time) parameters. Through the implemented smart contacts, top $K$ miners receive the highest votes and are selected as winners of each round. Consequently, these winners contribute their models to form an updated global model, and the implemented smart contracts fairly reward them based on the significance of their contributions. This system addresses critical challenges in conventional FL-based consensus mechanisms, such as fairness and incentive alignment, while enhancing overall efficiency. The contributions of this paper are presented as follows:

\begin{itemize}
  \item We propose a novel multi-winner consensus mechanism based on FL. In this framework, training is achieved globally through the contributions of all miners, improving the fairness in mining competition, as opposed to PoFL.
  \item We present a novel model evaluation mechanism based on miners' test data and the utilization of their local models. We also explore the robustness of this method by proposing and executing potential attacks, demonstrating their failure in the results section.
  \item Lastly, we propose a novel rewarding mechanism that considers the significance of the contribution of each winning miner in reward distribution. Through comprehensive experiments, we show that our reward system is fair both within and across all rounds.
\end{itemize}

\section{Related Works}
Given the multimodal nature of our proposed framework, we categorize related works into three distinct areas for thorough comparison: i) related consensus mechanisms, ii) studies on FL, and iii) the application of blockchain technology within FL contexts.

\subsection{Consensus Mechanisms}
Initially, Nakamoto \cite{nakamoto2017bitcoin} introduced Proof-of-Work (PoW) for Bitcoin. To maintain the ledger's integrity, miners must find a nonce that meets specific hash conditions when adding to the block header. The search for an appropriate nonce value is notably computation-intensive, driving research into more efficient alternatives, leading to solutions for resource-efficient and energy-recycling consensus mechanisms \cite{bada2021towards}.

Several studies have introduced consensus mechanisms that are more efficient and less energy-intensive than Proof-of-Work (PoW) \cite{lashkari2021consensus}. For instance, Proof-of-Stake (PoS), suggested by Nguyen et al. \cite{nguyen2019pos}, grants validators with high stakes the privilege to add new blocks. Proof-of-Work-Time (PoWT) \cite{vericonomy} seeks to increase mining efficiency and reduce computational waste by incorporating a block time parameter \cite{lashkari2021consensus}. Proof-of-Burn (PoB) \cite{karantias2020proof} encourages miners to ``burn" coins by sending them to an address from which they cannot be recovered in exchange for increased virtual mining power. Expanding on these innovations, Proof-of-History (PoH) \cite{yakovenko2018solana} provides a chronological verification of events, enhancing blockchain efficiency by confirming transaction sequences. Proof-of-Activity (PoA) \cite{Bentov2014poa} combines the principles of PoW and PoS, starting with a mining process and subsequently transitioning to a stake-based validator selection for block finalization, thereby bolstering both network security and energy efficiency. 

Further studies have considered replacing the PoW puzzle with productive tasks to conserve energy \cite{Qu2021pofl}. Proof-of-eXercise (PoX) \cite{Shoker2017pox} and Proof-of-Useful-Work \cite{dong2019proofware} repurpose mining efforts for real-world scientific computations and polynomial problem-solving, respectively. Primecoin \cite{king2013primecoin} searches for prime number sequences. Our focus is on consensus mechanisms that integrate learning algorithms into their core.

Bravo-Marquez et al \cite{BravoMarquez2019ProofofLearningAB} developed Proof-of-Learning (PoL), wherein ``trainers" submit machine learning models for evaluation by ``validators," and the most effective model receives rewards. PoFL \cite{Qu2021pofl} engages miners in training models within mining pools to compete for rewards, leading to an unfair advantage for the winning pool in early rounds as the victorious model is not shared among pools. Moreover, the network does not train a unified global FL model since each pool develops its own model architecture. Similarly, Proof-of-Training-Quality \cite{lu2020poq} faces similar issues by implementing FL in local committees.

In conclusion, we identify a gap in existing consensus mechanisms: a lack of support for a fair and secure, globally validated FL model. Our paper addresses this by proposing a novel mechanism to fill this gap.

\subsection{Federated Learning}
McMahan et al. \cite{mcmahan2017communication} originally proposed FL to train machine learning models across distributed networks. A standard FL network comprises a central server and several client devices. The process begins with the server distributing the global model to each client. Clients then train this model locally using their private datasets. Once training is complete, clients send their model updates back to the central server. The server aggregates these updates to enhance the global model. This cycle of distribution, local training, and aggregation constitutes one round of FL. Multiple such rounds are conducted, allowing the global model to progressively improve in accuracy and performance as it learns from a broader range of data across the network. While FL was first introduced within the realm of machine learning, numerous studies, including this paper, have explored its application in training deep learning models in a distributed manner \cite{hao2019towards, li2021deepfed, yin2020fdc, ferrag2021federated}.

\subsection{Blockchain-enabled federated learning}

Although FL provides significant privacy benefits, privacy still remains a primary concern when implementing this algorithm. In an FL network, no raw data is communicated; however, sharing model parameters can still pose privacy risks for clients \cite{li2020fl}. For example, the central authority may infer details about local training datasets by conducting Membership Inference Attacks (MIA) on the received model parameters \cite{shokri2017mia}. Efforts to address these privacy concerns, such as the integration of differential privacy into federated or deep learning models, often result in reduced model utility \cite{li2020fl}. Additionally, the standard FL algorithm lacks an incentive mechanism to motivate clients to contribute their computational resources honestly in training the global model \cite{zhu2023blf}. To address these challenges, Blockchain-enabled FL systems have emerged \cite{qu2022blockchain}. 


The decentralized management of blockchain systems, along with their ability to reward clients, substantially benefits FL. Originally, Kim et al. \cite{kim2020blockfl} proposed BlockFL, the first blockchain-enabled FL method that utilizes a decentralized ledger to exchange local model updates. Li et al. \cite{li2021bflc} propose a blockchain-based FL framework with committee consensus. In this framework, a committee of honest nodes is randomly selected to validate the model updates proposed by other trainers. Blockchain-empowered FL is utilized in various applications, including the industrial internet \cite{lu2020IoT, zhang2021IoT, Qu2021cognitive}, smart healthcare \cite{kumar2021ctimaging, rahman2020secure}, and wireless network infrastructure \cite{cui2022creat, pokhrel2020auto}, \cite{zhu2023blf}.

\section{Design}
In this paper, we extend the principles of FL to propose a consensus mechanism integrated with a collaborative deep learning algorithm. The traditional FL network comprises a central server and multiple clients. These clients train a shared global machine learning model using their private local datasets and subsequently upload their model updates to the central server. The server aggregates these updates to form an enhanced global model, completing what is known as a round. In this section, we introduce a series of steps that must be taken, in each round, to implement PoCL.

To effectively implement our FL-based consensus mechanism, we define the following entities within our network:

\emph{(i) Administrator}: An entity responsible for altering user-defined values including the number of rounds, number of winning miners, and deadlines for each step. The administrator is also responsible for notifying the miners about the details of each step. To ensure that this role does not compromise the network's decentralization, the Administrator's functions can be implemented through decentralized governance mechanisms. For instance, changes proposed by the Administrator could require approval through a distributed consensus process among selected stakeholders or be automated through smart contracts that execute based on predefined rules agreed upon by the network participants.

\emph{(ii) Requesters}: In this network, users who commit deep learning tasks are called requesters. Each request contains a deep learning model, which will be trained globally, and a list of publicly available datasets for training. Since sharing private data with multiple miners might bring about privacy concerns for the requesters, we limit our consideration for the training data of each model to only publicly available datasets. Requests are saved in a queue to select the global model trained at each round.

\emph{(iii) Miners}: In charge of adding new blocks to the blockchain, miners train the global deep learning models, predict test records of other miners, and vote on the predictions made by others. Miners compete with each other to be among the winning miners by training the most superior model in a short time. Winner miners are rewarded according to the significance of their contribution to the global models.

\begin{figure}[h]
  \centering
  \includegraphics[width=\linewidth]{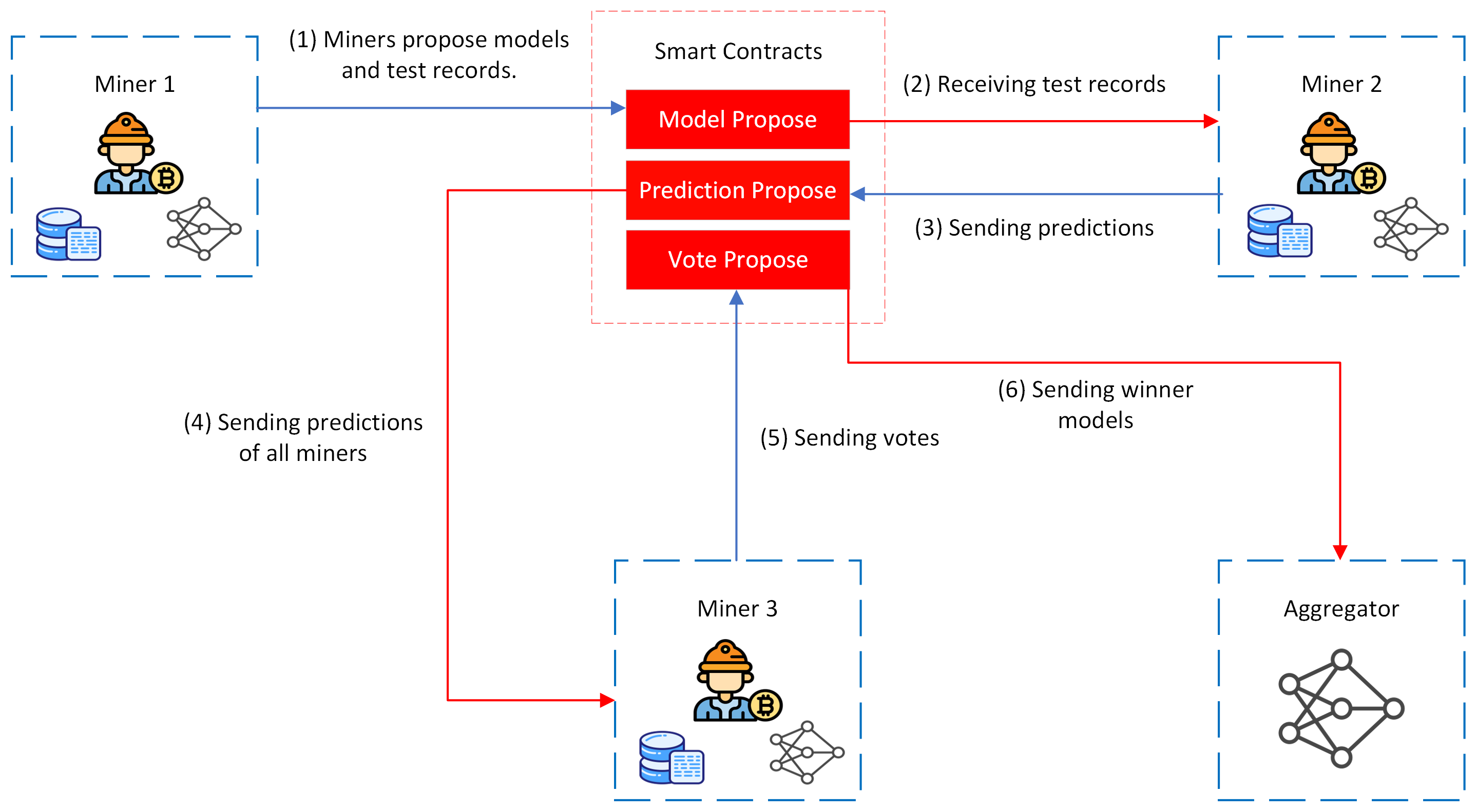}
  \caption{Proof-of-Collaborative-Learning (PoCL) Design.}
  \label{fig:design}
\end{figure}

\emph{(iv) Aggregator}: An off-chain program that supervises the aggregation of the winner models. The aggregator can shift between different variations of FL according to the requester’s preference. Furthermore, the program computes the contribution of each winning miner and reports it to the blockchain to reward them accordingly.

\emph{(v) Users}: Similar to any user in a blockchain network, they submit transactions to be mined and added to the blockchain. All submitted transactions are stored in a transaction pool, from which miners select transactions to mine. 

Furthermore, we assume that the peers within the blockchain network possess the capability to execute and store smart contracts. Given these assumptions, we propose a series of actions to be undertaken in each round to achieve a consensus validated by global FL. Figure~\ref{fig:design} and Algorithm~\ref{alg:design} illustrate the key steps of our proposed framework.


\subsection{Model Proposal}
Each round is started by selecting a global deep-learning model from the request queue and sending it to the miners, who in turn ask for validated transactions to mine and start training. After receiving their designated transactions and training the global model using their local data, the miners form a model proposal block including the hash of the trained model and a set of unlabeled test records. A curated set of test cases is distributed to clients, ensuring that these cases are representative of the overall data distribution while not compromising data privacy. This step is crucial for evaluating the local models under similar conditions, providing a fair basis for comparison.

At the initiation of each round, miners are notified of a period called the model proposal deadline. Miners should take the actions mentioned above and send their model proposal block before the deadline passes; otherwise, they will not be included in the following steps. The administrator determines the model proposal deadline which can vary from one system to another based on distinct miner needs and global models. In some cases, a global model may require more training time to reach an adequate training level. Hence, we consider this as a custom value for the employers of our system.

\subsection{Prediction Proposal}
To select the best models, an evaluation step must be performed. In our FL framework, We address the quality of the locally trained model by implementing a distributed evaluation protocol. This protocol allows us to assess the performance of local models trained on miners' devices, ensuring that contributions to the global model are made by the most effective miners.

After passing the model proposal deadline, the submitted test records are collected and sent to miners who participated in the model proposal step. The miners predict the given test records by feeding the inputs to the model. Afterward, they forward the predictions to a smart contract, managing all miner predictions. 

To prevent miners from deducing the correct labels of the test records through any means other than their trained models, we restrict the prediction proposal phase by setting a deadline. This deadline is carefully calibrated to be just sufficient for the execution of a forward pass of the global model. Proposals submitted after this deadline are not accepted to ensure the integrity of the evaluation process.

Setting the prediction proposal deadline is another responsibility of the administrator, and its duration may vary between systems in distinct cases. Nevertheless, the general rule is to conduct an experiment and calculate the average time it takes miners to complete a forward pass of the global model and send the results. This empirically determined duration can then serve as the basis for establishing the prediction proposal deadline.

\subsection{Vote Proposal}
The vote proposal phase is crucial in identifying and distinguishing between honest miners and potential adversaries. During this phase, miners evaluate the predictions submitted by their peers in the preceding step, considering two primary criteria: the loss value associated with the predictions and the time required for the miners to submit their predictions. Firstly, they rank predictions from most to least accurate. In instances of identical accuracy, preference is given to submissions that were made earlier. This prioritization encourages miners to comply with the guidelines set for the prediction proposal phase and to submit their predictions promptly. Furthermore, this approach enhances the system's resilience against potential attacks discussed in the Security Concerns section.

\subsection{Winner Selection}
A Chaincode collects all submitted votes and selects the top \emph{K} miners with the highest number of votes as winners of the round. Subsequently, their trained models are transmitted to the aggregator to perform a FL combination algorithm, such as Federated Averaging (FedAvg) \cite{mcmahan2017communication}. 
\begin{algorithm}
\caption{The general workflow of the framework}
\label{alg:design}
  \Class{Miner:}{
      \Comment{"blc" is the blockchain containing all Chaincodes.}
      \Function{mine():}{ 
        \var{miningTrx} $\assign$ \FuncCall{blc.getTrx}{self.id}\;
        \var{testRecords} $\assign$ \FuncCall{getTestRecords}{}\;
        \FuncCall{model.fit}{localData}\;
        \var{modelHash} $\assign$ \FuncCall{md5}{model}\;
        \FuncCall{blc.submitModel}{modelHash, testRecords}\;
      }
      \Function{predict():}{
        \var{testRecords} $\assign$ \FuncCall{blc.getTestRecords}{}\;
        \var{preds} $\assign$ \FuncCall{model}{testRecords}\;
        \FuncCall{blc.submitPreds}{preds}\;
      }
      \Function{vote():}{
      \Comment{"times" are the times of predictions recorded by the Chaincode}
        \var{preds}, \var{times} $\assign$ \FuncCall{blc.getPreds}{self.id}\;
        \var{losses} $\assign$ \FuncCall{lossFn}{preds, testLabels}\;
        \var{votes} $\assign$ \FuncCall{sort}{losses, times}\;
        \FuncCall{blc.submitVotes}{votes}\;
      }
  }
  
  \Class{Blockchain:}{
  \Comment{"agg" is the aggregator.}
      \Function{start():}{
        \For{miner \In{in} miners:}{
            \FuncCall{notify}{miner, "mine"}\;
        }
        \FuncCall{setTimeout}{getPreds, modelDeadline}\;
      }
      \Function{getPreds():}{
        \For{miner \In{in} miners:}{
            \FuncCall{notify}{miner, "predict"}\;
        }
        \FuncCall{setTimeout}{getVotes, predDeadline}\;
      }
      \Function{getVotes():}{
        \For{miner \In{in} miners:}{
            \FuncCall{notify}{miner, "vote"}\;
        }
        \FuncCall{setTimeout}{selectWinners, voteDeadline}\;
      }
      \Function{selectWinners():}{
        \var{votes} $\assign$ \FuncCall{gatherAllVotes}{}\;
        \var{winners} $\assign$ \var{votes[:k]}\;
        \FuncCall{runTrx}{winners}\;
        \var{rewards} $\assign$ \FuncCall{agg.combine}{winners}\;
        \FuncCall{giveRewards}{winners, rewards}\;
      }
  }
\end{algorithm}
The aggregator verifies the integrity of each winning model by comparing its hash against the proposed hash from the initial step to ensure that the model has remained untampered since its proposal. 

\subsection{Block Creation}
In the \emph{Model Proposal} step, each miner is assigned some validated transactions to mine. In this step, A smart contract gathers the designated transactions of the winner miners into a single block and appends it to the ledger.

\begin{figure}[h]
  \centering
  \includegraphics[width=\linewidth]{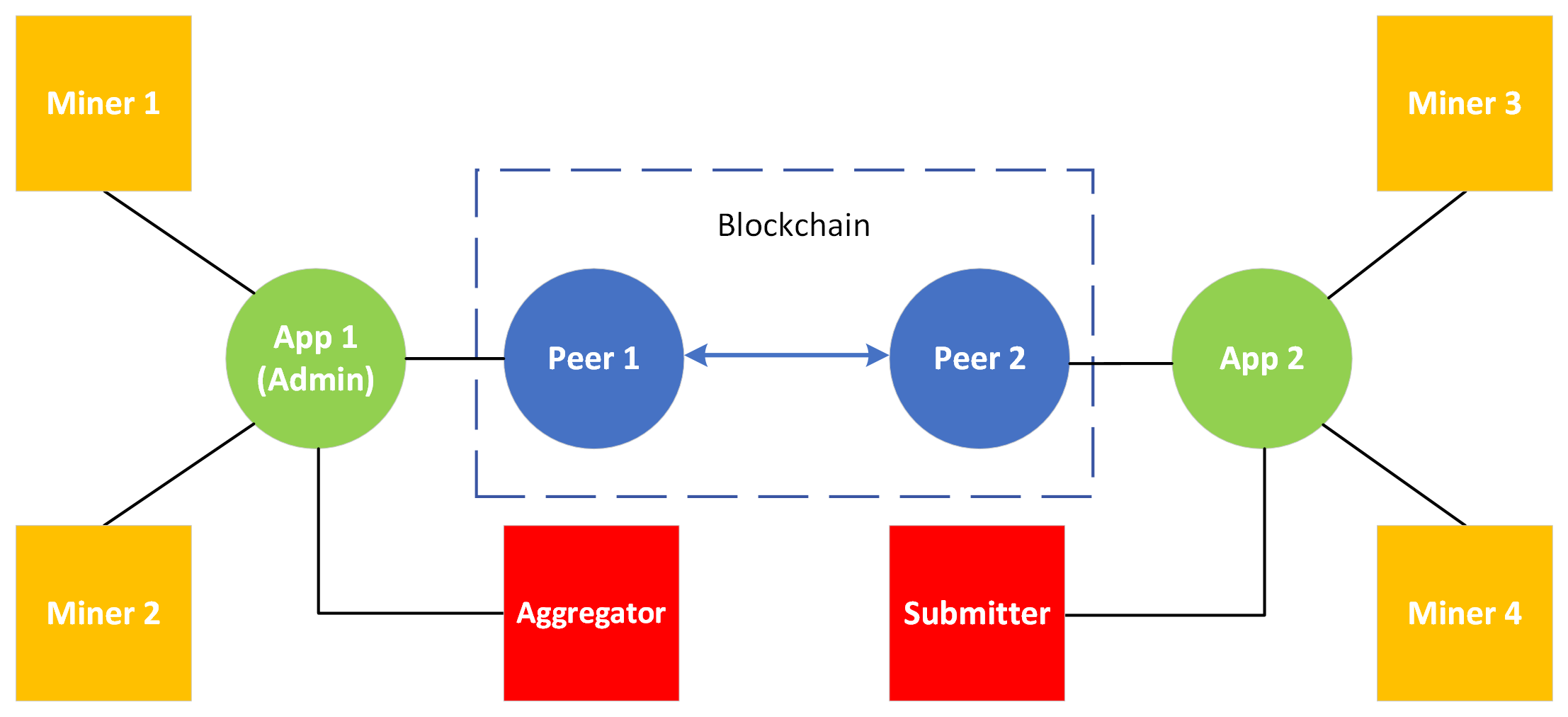}
  \caption{Network architecture}
  \label{fig:implementation}
\end{figure}

\begin{figure}[h]
  \centering
  \includegraphics[width=\linewidth]{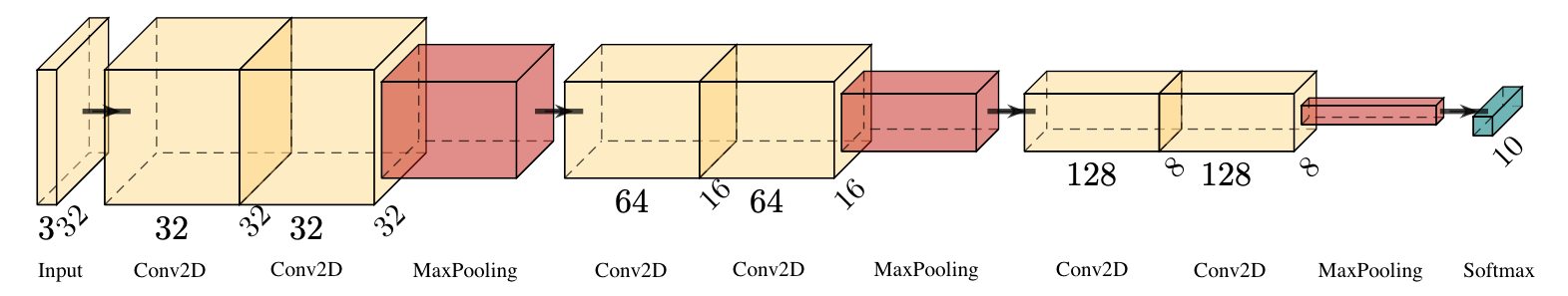}
  \caption{The global model architecture.}
  \label{fig:model}
\end{figure}

\subsection{Reward Mechanism}
In this framework, we introduce a fair reward distribution mechanism designed to compensate winning miners in proportion to the significance of their contributions to the global model in the preceding round. We denote the global model as $M_{G} = [\tilde{W^{1}}, \tilde{W^{2}}, ..., \tilde{W^{L}}]$ and the local model of the i-th miner as $M_i$ where $M_i = [W^{1}, W^{2}, ..., W^{L}]$. In these lists, $W^{1}$ and $\tilde{W^{1}}$ are the weights of the first layers of the local and global models, respectively, and $L$ is the number of layers in the model. Given that the layer sizes can vary, the reward for winning miner $i$ is calculated based on a formula that accounts for these variations:

\begin{equation}
R_{i} = \frac{1}{L} \sum_{l} \frac{1}{N_l} \sum_{n} |W_{n}^{l} - \tilde{W_{n}^{l}}|
\end{equation}

In simpler terms, we calculate the average impact that each winning miner has on the current global model and base their rewards on this calculation. At the end of each round, the aggregator sends this information to the smart contract that issues rewards to the winning miners. In the results section, we show how this reward mechanism enhances the fairness of our proposed framework.

 

\section{Security Concerns}
A potential threat to our FL validated consensus mechanism is the risk of miners not following the training rules. Specifically, there is a concern that some miners might attempt to mislead others by generating predictions using other supervised or unsupervised learning algorithms, such as K Nearest Neighbors (KNN), rather than adequately training the global FL model. In this section, we identify two scenarios where an adversary might employ such techniques, and we argue that in both scenarios, such adversaries will not succeed. Additionally, in the results section, we substantiate our assertion by conducting comprehensive simulations of KNN-based attacks across varying data sizes.

Firstly, we define the dataset $\mathcal{T}_{train} = \{x_i\}_{i=1}^{N_{train}}$ where $x_i \in \mathbb{R}^D$ as our training dataset.  
\begin{figure}[h]
  \centering
  \includegraphics[width=\linewidth]{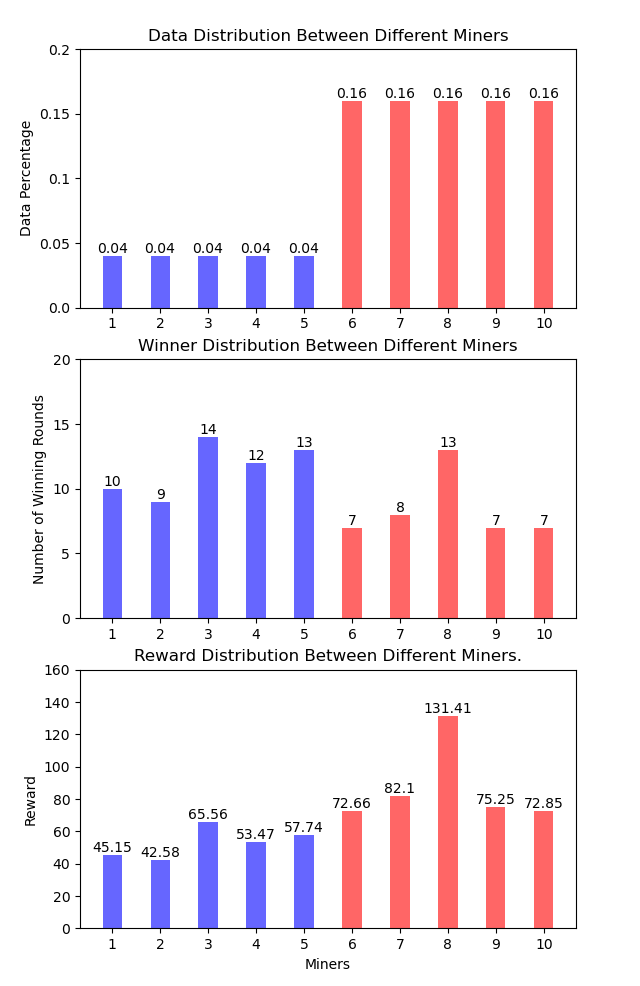}
  \caption{The relationship between data size, rewards, and the number of winning rounds.}
  \label{fig:relationship}
\end{figure}
If $D$ is considerably large, particularly when $\mathcal{T}_{train}$ comprises image data, alternative methods to deep learning models generally do not produce results competitive with those achieved by miners training a global deep learning model. Therefore, an adversary employing simpler machine learning algorithms in an attempt to mislead other miners will fail in this competitive scenario.


However, the effectiveness of K-Nearest Neighbors (KNN) algorithms can be similar to deep neural networks when data has low dimensionality. That is to say, either approach achieves satisfactory and similar results. In such instances, the complexity introduced by deep convolutional neural networks (CNNs) may not significantly outperform simpler Fully Connected (FC) models. Therefore, we consider a simple, fully connected neural network for our global model in scenarios characterized by low-dimensional data. This decision is based on the principle that the data's simplicity does not require the advanced feature extraction of CNNs, which are suited for high-dimensional datasets like images.

Assuming that $N_{train}$ and $N_{test}$ are the test and training data sizes, respectively, the time complexity of the KNN algorithm is:

\begin{equation}
T(N_{train}, N_{test}, D) = O(N_{test}.N_{train}.D + N_{test}.\log(N_{test}))
\end{equation}

\begin{figure}[h]
  \centering
  \includegraphics[width=\linewidth]{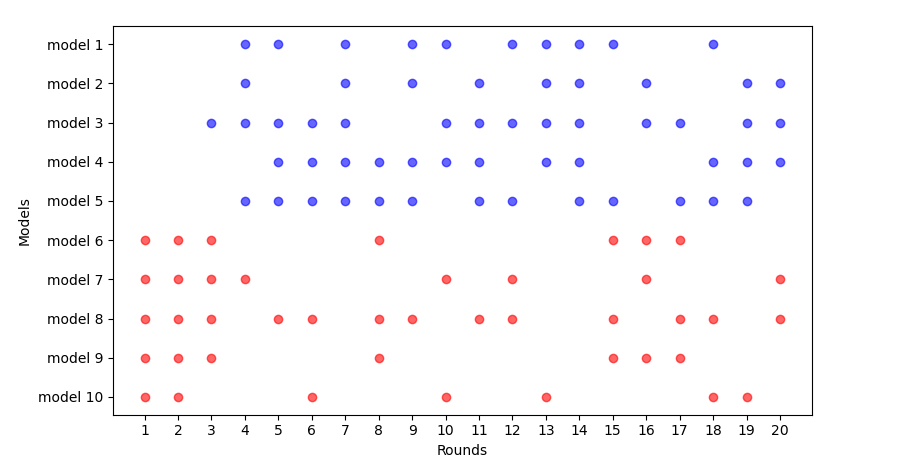}
  \caption{The winning miners for each mining round.}
  \label{fig:winner-rounds}
\end{figure}

where $O(N_{test}.N_{train}.D)$ is the time complexity of computing the distance between the test records and all the training records and $N_{test}.\log(N_{test})$ is for sorting the distances. Similarly, the time complexity of a forward pass in a fully connected neural network is:

\begin{equation}
T(N_{test}, D, H_{max}) = O(N_{test}.D.H_{max})
\end{equation}

Where $H_{max}$ is the maximum layer size in the network. This equation is derived based on the most computationally intensive matrix multiplication in the forward pass. Considering the above equations, unless for extremely large values of $H_{max}$, the time complexity of a forward pass is smaller than that of a KNN algorithm for a low-dimensional dataset $\mathcal{T}$. Consequently, in the proposed voting step, the deep learning model is favored for faster computation when these algorithms perform equally well.

\section{Implementation}
We utilized Hyperledger Fabric \cite{androulaki2018hyperledger}, a modular and permissioned blockchain platform, to implement PoCL\footnote{https://github.com/tcdt-lab/FL-Validated-Learning}. The platform's Chaincode-centric architecture, where Chaincodes function as smart contracts, allowed us to seamlessly integrate our consensus mechanism as a network layer on the existing Hyperledger Fabric infrastructure. We developed smart contracts (Chaincodes) for each critical operation within our comprehensive design. This design ensures that no single miner or central authority oversees network operations; instead, governance is distributed across a suite of smart contracts, ensuring decentralized control. Figure~\ref{fig:implementation} shows a high-level view of our implemented network and demonstrates how distinct components of the PoCL framework are connected.

\begin{figure}[h]
  \centering
  \includegraphics[width=\linewidth]{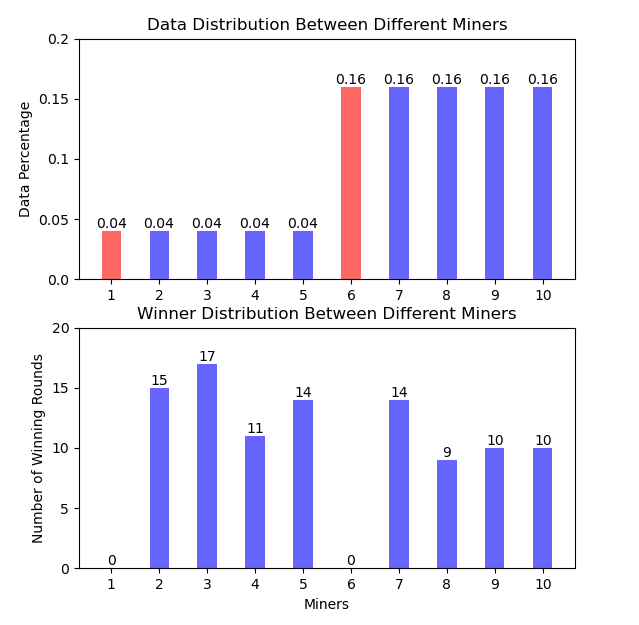}
  \caption{The result of KNN attacks taken by adversaries with both small (miner 1) and large (miner 6) datasets.}
  \label{fig:knn}
\end{figure}

Our implementation revolves around the functionality of the following elements:

\emph{(i) Peers}:  As essential nodes in Hyperledger Fabric networks, peers maintain a copy of the ledger and are responsible for endorsing transactions by executing Chaincodes (smart contracts). They validate transaction outputs against the current ledger state before committing them to the blockchain, ensuring consistency and integrity. 

\emph{(ii) Applications}: Applications serve as gateways for communicating with the peers of the Hyperledger Fabric network. They enable the submission of transactions and the triggering of Chaincode functions. In our implementation, each peer is associated with one application to facilitate communication. Additionally, the administrator is implemented as one of these applications to simplify the contact and notification processes required for interaction with miners. 

\emph{(iii) Miners}: Miners are implemented as Flask servers waiting for applications to notify them to start training. They predict test records and vote on the predictions made by other miners. The TensorFlow framework performs this training process.

\emph{(iv) Aggregator}: Similar to the miners, the aggregator is implemented as a Flask server that awaits the command to aggregate updates into a new global model using the FedAvg algorithm. Notably, our design is flexible and can be adapted to incorporate other variations of FL.

\begin{figure*}[t]
  \centering
  \includegraphics[width=\linewidth]{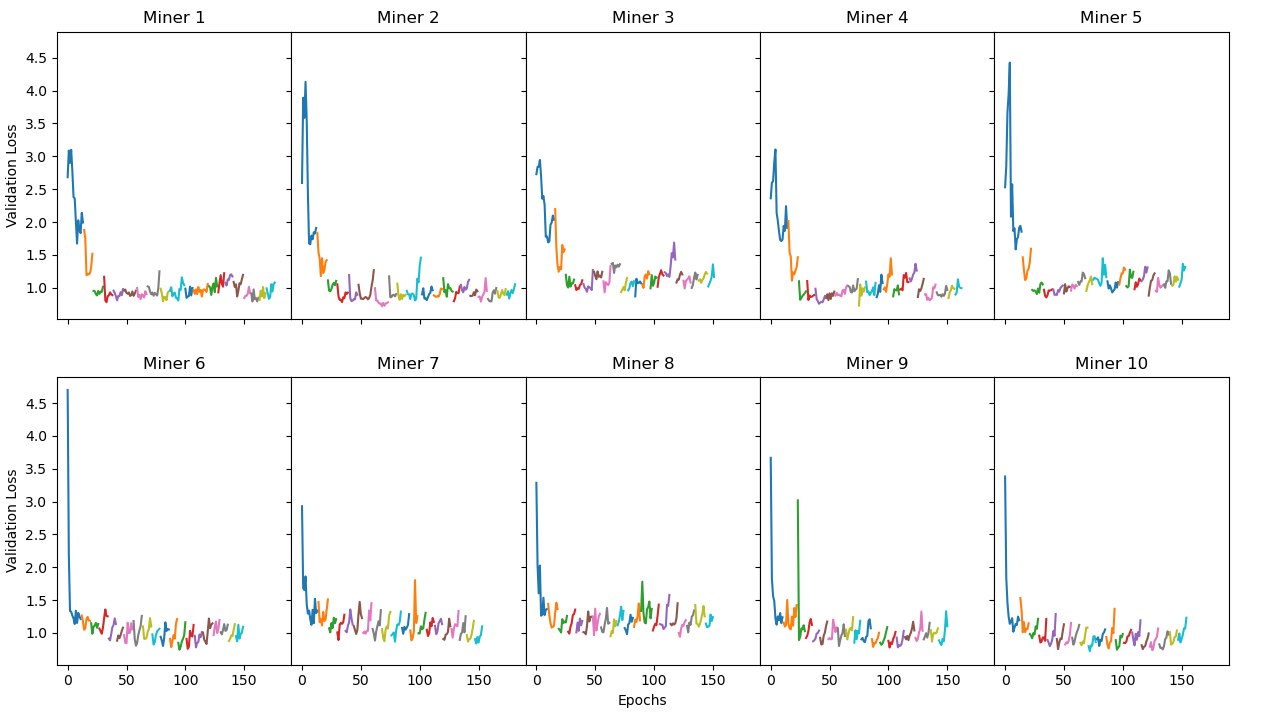}
  \caption{Validation loss through 20 mining rounds. In each subplot, different colors indicate different rounds of training.}
  \label{fig:val-loss}
\end{figure*}

\emph{(v) Submitter}: To accurately simulate users' activities on the blockchain, a program is built to submit transactions to the blockchain at a custom rate. The task of our blockchain is to supervise a cryptocurrency system; therefore, the submitter proposes transfer transactions from one wallet to another.

\emph{(vi) Adversaries}: The miners who attempt to invade the privacy of the system by conducting three malicious actions. First, instead of training the global model using their local datasets, they replace every trained parameter with zero to ruin the performance of the global model should they win the current round. In addition, instead of computing a forward pass to make predictions, they utilize a KNN algorithm. At last, while every honest miner votes from best to worst predictions, they vote from worst to best to increase the chance of winning for less efficient models. 

\section{Results}
We trained a deep Convolutional Neural Network \cite{aloysius2017review}, as depicted in Fig. \ref{fig:model}, to classify the CIFAR-10 dataset using 10 miners with identical capabilities. In each round, the top $K=5$ miners are selected based on the loss value and prediction time, and their models are aggregated using the FedAvg algorithm. To assess the fairness of our framework, we unevenly distributed the dataset among miners, with miners six to ten receiving four times as many data records as miners one to five. This experimental setup was chosen to evaluate the impact of data volume on each miner's chance of winning in competitive rounds.

Fig. \ref{fig:val-loss} shows the validation loss of each local model over 20 rounds of training. As demonstrated by these plots, all local models show improvement over the course of training.

Furthermore, to assess the relationship between data size and competitive success, we illustrate the total number of winning rounds for each miner in Figure \ref{fig:relationship}. Counterintuitively, miners with smaller data sizes win more rounds. These unexpected results necessitated further examination of the distribution of winning miners across each round, as shown in Figure \ref{fig:winner-rounds}. This illustrates that miners with larger data sizes are more likely to win initial rounds. Additionally, from Figure \ref{fig:val-loss}, we infer that their contributions to the global model in these rounds are highly significant, evidenced by the considerable decrease in validation losses and increase in validation accuracy. Our reward mechanism appropriately acknowledges the contributions of these miners by assigning reward values, as illustrated in Figure \ref{fig:relationship}. Although miners with smaller data sizes win more rounds, they receive fewer rewards than miners with larger data sizes. This analysis confirms that our framework promotes fair competition in each round and ensures an equitable reward distribution among winning miners across all rounds. 

Moreover, we conducted a similar experiment using the same hyperparameters to assess the impact of the KNN attacks on the system. In this experiment, we introduced two adversaries, miners one and six, with the latter possessing four times more data records than the former. The results of this setting are demonstrated in Figure \ref{fig:knn}, showing that neither adversary won any competition round, regardless of data size.




\section{Conclusion}
In this paper, we proposed Proof-of-Collaborative-Learning (PoCL), a multi-winner FL validated consensus mechanism that recycles the energy of the original proof of work. This framework trains the requested models globally using the computation power of the contributing miners. To evaluate miners' performance and select winners, we proposed a novel evaluation step that relies on the predictions miners make on each other's test records. In addition, we verified the robustness of our proposed evaluation mechanism by simulating an attack scenario on the mechanism and demonstrated how these attacks are ineffective in compromising the system. Finally, we proposed a novel reward distribution mechanism that compensates winning miners according to the significance of their contributions. Through extensive experiments, we demonstrated that our reward distribution algorithm fairly compensates miners both within and across all rounds. 

\section{Future Works}
In this paper, we proposed a general framework for achieving FL validated consensus through the global contribution of miners. We purposefully designed this framework to be expandable in many directions, including:

\begin{enumerate}
  \item In this study, we assumed the training data to be publicly available to prevent security concerns of requesters. Nevertheless, a privacy data-sharing approach can be utilized to train the global model using private data. This approach is particularly useful in private blockchains due to the provided transparency.
  \item For simplicity, we assumed that the requesters select the number of training rounds for their proposed models. However, one might not be aware of the correct number of training rounds required for training one's global model to convergence. In this setting, the framework can be adjusted to stop global training using approaches such as early stopping or by calculating the difference between the global models of consecutive rounds.
\end{enumerate}

\bibliography{ref}
\bibliographystyle{IEEEtran}

\end{document}